\documentclass[twocolumn,showpacs,pre,UTF8]{revtex4}
\usepackage{amsmath}
\usepackage{txfonts}
\usepackage[titletoc]{appendix}
\usepackage{amssymb}
\usepackage{array}
\usepackage{mathrsfs}
\usepackage{subfigure,overpic}
\usepackage{dcolumn}
\usepackage{graphicx}
\usepackage{epstopdf}
\usepackage[colorlinks,
            linkcolor=blue,
            anchorcolor=blue,
            citecolor=blue]{hyperref}

\begin{document}

\title{Non-linear antidamping spin-orbit torque originating from intra-band transport on the warped surface of a topological insulator}
\author{Yong-Long Zhou$^{1}$}
\author{Hou-Jian Duan$^{1,2}$}
\email{dhjphd@163.com}
\author{Yong-jia Wu$^{1}$}
\author{Ming-Xun Deng$^{1,2}$}
\author{Lan Wang$^{4}$}
\author{Dimitrie Culcer$^{3}$}
\author{Rui-Qiang Wang$^{1,2}$}
\email{wangruiqiang@m.scnu.edu.cn}
\affiliation{$^{1}$Guangdong Provincial Key Laboratory of Quantum Engineering and Quantum Materials, School of Physics and Telecommunication Engineering, South China
Normal University, Guangzhou 510006, China}
\affiliation{$^{2}$Guangdong-Hong Kong Joint Laboratory of Quantum Matter, Frontier Research Institute for Physics, South China Normal University, Guangzhou 510006, China}
\affiliation{$^{3}$School of Physics, The University of New South Wales, Sydney 2052, Australia}
\affiliation{$^{4}$School of Science and ARC Centre of Excellence in Future Low-Energy Electronics Technologies, RMIT Node, RMIT University, Melbourne, VIC 3000, Australia}

\begin{abstract}
Motivated by recent experiments observing a large antidamping spin-orbit torque (SOT) on the surface of a three-dimensional topological insulator, we investigate the origin of the current-induced SOT beyond linear-response theory. We find that a strong antidamping SOT arises from intraband transitions in non-linear response, and does not require interband transitions as is the case in linear transport mechanisms. The joint effect of warping and an in-plane magnetization
generates a non-linear antidamping SOT which can exceed the intrinsic one by several orders of magnitude, depending on warping parameter and the position of Fermi energy, and exhibits a complex dependence on the azimuthal angle of the magnetization. This nonlinear SOT provides an alternative explanation of the observed giant SOT in recent experiments.
\end{abstract}

\maketitle

Electrical control of magnetic systems has a strong potential for technological applications such as fast magnetic-based storage and computational devices \cite{I.Zutic2004}. Recent works in this fast-evolving field
have demonstrated that large spin-orbit coupling in ferromagnet/heavy-metal
(FM/HM) bilayers can produce strong enough spin-orbit torques (SOTs) to switch the magnetization in the overlayer. Compared to conventional spin transfer torques in ferromagnet/insulator/ferrometal
biheterostructures\cite{G.Prenat2015,A.Manchon2018}, this SOT has a lower
current- and energy-threshold required for magnetization switching%
\cite{G.Prenat2016,K.-S.Lee2016}. In these systems, the antidamping-like (ADL)
torque has the same form as the Gilbert damping term in the
Landau-Lifshitz-Gilbert equation\cite{T.Yokoyama2011} but has the opposite
sign, and competes against Gilbert damping to switch the magnetization.
Therefore, a large ADL torque is of particular importance for increasing the
efficiency of magnetization switching. Antidamping torques in these structures arise from either the spin Hall effect (SHE) within the bulk of heavy metals\cite%
{L.Liu2012a,L.Liu2012b,L.Liu2012c,J.Sinova2015,A.Manchon2018,J.Sinova2017}
or the Rashba-Edelstein effect (or the inverse spin galvanic effect) at
inversion-symmetry broken interfaces\cite%
{M.Trushin2007,I.M.Miron2011b,T.D.Skinner2015,L.Chen2018}. They may also stem from the intrinsic Berry curvature \cite%
{H.Kurebayashi2014}, without being related to a bulk SHE.
\par
Besides heavy metals, topological insulators \cite
{M.H.Hasan2010,X.L.Qi2011}, in which the intrinsic strong spin-orbit
coupling is large enough to invert the band structure, are the most promising candidates towards efficient transfer of angular momentum between the charge current and the local magnetization. Recent experiments
in FM/TI layered structure reported a giant SOT\cite%
{Y.Fan2014,Y.Wang2017,M.DC2018,A.R.Mellnik2014,Y.Wang2015,J.Han2017,Y.Fan2016}
even at room temperature. Compared to FM/HM systems the current density required for
magnetization switching\cite{Y.Wang2017,M.DC2018,J.Han2017,C.H.Li2016} in FM/TI bilayers
is one to two orders of magnitude smaller, and the corresponding effective spin Hall angle%
\cite{Y.Fan2014,J.Han2017} is several times larger. Most experiments confirm that the giant
SOT originates from the surface states, e.g., the charge-to-spin current conversion efficiency increases when the Fermi energy is within the TI bulk gap rather than in the bulk states \cite{M.Mogi2021}, excluding
contributions from the SHE and Rashba-Edelstein effect. In this context, understanding the origin of the large ADL-SOT in FM/TI bilayers becomes a crucial issue.
\par
Theoretically, there have been many efforts to explain the emergence of large
SOTs, especially the antidamping component, at the magnetic surfaces of
topological insulators using linear response theory. Garate and Franz%
\cite{I.Garate2010,kur} ascribed the SOTs in FM/TI bilayers to a
topological magnetoelectric effect with emphasis on its dissipationless Hall
current for Fermi energies in the Dirac gap.
Extending it to finite Fermi energies, this dissipationless
damping was also found to arise from intrinsic inter-band transitions%
\cite{P.B.Ndiaye2017,T.Chiba2017,A.Sakai2014,T.Yokoyama2010}. Mahfouzi et al.%
\cite{F.Mahfouzi2015} obtained an antidamping torque by considering spin-flip reflection at an interface. Nevertheless major questions remain unanswered. Theoretically\cite%
{I.Garate2010,P.B.Ndiaye2017,T.Chiba2017,A.Sakai2014,T.Yokoyama2010,T.Chiba2020,T.Gao2018,kur},
the ADL-SOT due to the TI surface states has been expressed in the general form $\mathbf{\tau }_{D}=\tau _{d}m_{z}\mathbf{m}\times e \mathbf{E}$ where $\mathbf{m}$ is a unit magnetization vector and $\mathbf{E}
$ is the electric field. This form does not explain experimental observations: the ADL-SOT is quite weak and vanishes if $m_{z}=0$, and the in-plane magnetization $m_{x/y}$ has no effect on the SOT strength $\tau _{d}$. Nevertheless, in many recent experiments on FM/TI bilayers\cite%
{A.R.Mellnik2014,M.DC2018}, a strong angular dependence of SOTs on the azimuthal angular or $m_{x/y}$ was widely observed even in the absence of $m_{z}$. Theoretically, it was even doubted that the experimental measurement method relying on the second harmonic Hall voltage could accurately determine the SOT due to the disturbance from asymmetric magnon scattering\cite{K.Yasuda2016}.
\par
In this Letter, we propose a mechanism for the generation of the ADL-SOT in the
nonlinear response regime, purely based on the topological surface states
with hexagonal warping, which is strong in realistic TIs\cite%
{R.S.Akzyanov2018,R.S.Akzyanov2018b}. Our work stands in sharp contrast to existing theories, which are exclusively based on linear response. Intriguingly, we find that the nonlinear spin polarization can produce a large ADL-SOT, caused by the interplay between the warping effect and the in-plane magnetization, which is known to have strong observable features in charge transport\cite{P.He2018,vig,bha}.
This non-linear phenomenon is distinguished from previous mechanisms and explains the features of the ADL-SOT observed experimentally.
\par
\emph{Theory for SOT} - The SOT exerted on the FM layer has the form
$\mathbf{\tau =}\frac{2J}{\hbar }\mathbf{m}\times \mathbf{S}$ with the spin
polarization $\mathbf{S=}\sum_{\chi }\frac{d^{d}\mathbf{k}}{(2\pi )^{d}}%
\mathbf{s}_{\chi }(\mathbf{k})f(\varepsilon _{\mathbf{k}}^{\chi }),$ where $%
J $ is the $s$-$d$ exchange energy, the superscript $d$ represents the
dimension, and $\mathbf{s}_{\chi }(\mathbf{k})=(\hbar /2)\langle \Psi _{%
\mathbf{k}}^{\chi }|\boldsymbol{\sigma }|\Psi _{\mathbf{k}}^{\chi }\rangle $
is the spin expectation in the $\chi $-th band with eigenvector $\Psi _{\mathbf{%
k}}^{\chi }$ and eigenvalue $\varepsilon _{\mathbf{k}}^{\chi }$. In the
absence of applied current, the distribution function $f(\varepsilon _{%
\mathbf{k}}^{\chi })$ is the Fermi-Dirac distribution function $%
f(\varepsilon _{\mathbf{k}}^{\chi })=f^{(0)}(\varepsilon _{\mathbf{k}}^{\chi
})=[e^{(\varepsilon _{\mathbf{k}}^{\chi }-\varepsilon _{F})/k_{B}T}+1]^{-1}$
with Fermi energy $\varepsilon _{F}$ and temperature $T$, and thus $\mathbf{S}
$ vanishes due to $\mathbf{s}_{\chi }(-\mathbf{k})=-\mathbf{s}_{\chi }(%
\mathbf{k})$ for spin-momentum locked surface states of TIs. When an
in-plane current is applied, the spin polarization $\mathbf{S=S}^{oc}\mathbf{%
+S}^{in}$ can arise from the two types of change. One originates from the change of the electron occupation $\delta f(\varepsilon _{\mathbf{k}}^{\chi
})=f(\varepsilon _{\mathbf{k}}^{\chi })-f^{(0)}(\varepsilon _{\mathbf{k}%
}^{\chi })$ within intraband due to acceleration by an electric field,
calculated by $\mathbf{S}^{oc}\mathbf{=}\sum_{\chi }\frac{d^{d}\mathbf{k}}{%
(2\pi )^{d}}\mathbf{s}_{\chi }(\mathbf{k})\delta f(\varepsilon _{\mathbf{k}%
}^{\chi })$. The other stems from the modification of the quasiparticle wave
functions \cite{X.Cong2017,H.Kurebayashi2014,I.Garate2009}, $\mathbf{S}%
^{in}=\sum_{\chi }\frac{d^{d}\mathbf{k}}{(2\pi )^{d}}\delta \mathbf{s}_{\chi
}(\mathbf{k})f(\varepsilon _{\mathbf{k}}^{\chi })$, where $\delta \mathbf{s}%
_{\chi }(\mathbf{k})=(\hbar/2) ${\rm Re}$\langle \Psi _{\mathbf{k}}^{\chi }|%
\boldsymbol{\sigma }|\delta \Psi _{\mathbf{k}}^{\chi }\rangle $ can be
traced to the interband contributions in analogy to the intrinsic
contribution to the anomalous Hall effect.
\par
We first discuss $\mathbf{S}^{oc}$ by employing the single-band
steady-state Boltzmann equation\cite{C.X.Liu2010},
\begin{equation}
-\frac{e}{\hbar }\mathbf{E}\cdot \triangledown _{\mathbf{k}}f(\varepsilon _{%
\mathbf{k}}^{\chi })=-\frac{f(\varepsilon _{\mathbf{k}}^{\chi
})-f^{(0)}(\varepsilon _{\mathbf{k}}^{\chi })}{\gamma (\mathbf{k})}.
\label{eq_f}
\end{equation}%
Here, we use the relaxation time approximation $\gamma (\mathbf{k})=\gamma $%
. Expanding $f(\varepsilon _{\mathbf{k}}^{\chi })=f^{(0)}(\varepsilon _{%
\mathbf{k}}^{\chi })+f^{(1)}(\varepsilon _{\mathbf{k}}^{\chi
})+f^{(2)}(\varepsilon _{\mathbf{k}}^{\chi })+...$ with $f^{(n)}(\varepsilon
_{\mathbf{k}}^{\chi })\propto $ $\mathbf{E}^{n}$ and then substituting it to
the above Boltzmann equation, one can find the recursive
relations for $n$-th order non-equilibrium distribution function,
\begin{equation}
f^{(n)}(\varepsilon _{\mathbf{k}}^{\chi })=\frac{e\gamma }{\hbar }\mathbf{E}%
\cdot \frac{\partial f^{(n-1)}(\varepsilon _{\mathbf{k}}^{\chi })}{\partial
\mathbf{k}}.  \label{eq_f1}
\end{equation}

\begin{figure}[t]
\centering \includegraphics[width=0.48\textwidth]{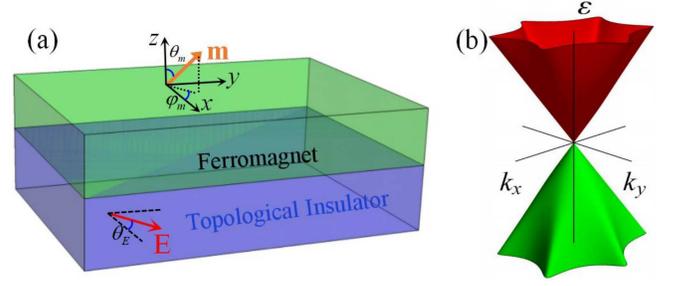}
\caption{(a) The FM/TI layered structure, where the orange arrow represents
the local magnetic moment with magnetization $\mathbf{m}$ in the FM layer, and the driven electric
field $\mathbf{E}=(E_{x},E_{y})=|\mathbf{E}|\left[\cos(\theta_E),\sin(\theta_E)\right]$ is applied in the TI layer. (b) Schematics of the band
structure for the warped surface states of TIs.}
\label{fig1}
\end{figure}

\emph{Nonlinear SOT from intra-band transitions} - We take a FM/TI heterostructure,
as shown in Fig. \ref{fig1}, as a sample system exhibiting a spin polarization in
response to an applied electric field. On the surface of a three-dimensional
TI, the effective Hamiltonian \cite{Y.L.Chen2009,L.Fu2009} reads
\begin{equation}
H_{TI}=\hbar v_{F}(\sigma _{x}k_{y}-\sigma _{y}k_{x})+\frac{\lambda }{2}%
(k_{+}^{3}+k_{-}^{3})\sigma _{z}+J\mathbf{m}\cdot \boldsymbol{\sigma },
\label{eq_H}
\end{equation}%
where $v_{F}$ is the Fermi velocity, $\boldsymbol{\sigma }=(\sigma
_{x},\sigma _{y},\sigma _{z})$ is the vector of Pauli matrice acting on real
spin, and $k_{\pm }=k_{x}\pm ik_{y}$ with $\mathbf{k}$ being the wave
vector. The first term is the Rashba-type spin-orbit coupling, the cubic-in-%
\textbf{k} term represents the hexagonal warping effect\cite%
{R.S.Akzyanov2018,R.S.Akzyanov2018b} of TIs with the warping parameter $%
\lambda $, and the FM layer is characterized by a local magnetization $%
\mathbf{m}=(m_{x},m_{y},m_{z})=[\sin (\theta_m) \cos (\varphi_m) ,\sin (\theta_m) \sin (\varphi_m) ,\cos
(\theta_m) ]$. The energy dispersion of the Hamiltonian in
Eq. (\ref{eq_H}) reads
\begin{equation}
\varepsilon ^{\chi }_{\mathbf{k}}=\chi \hbar v_{F}\sqrt{(k_{x}-Jm_{y}/\hbar
v_{F})^{2}+(k_{y}+Jm_{x}/\hbar v_{F})^{2}+\Lambda _{\mathbf{k}}^{2}},
\label{eq_Ek}
\end{equation}%
where $\Lambda _{\mathbf{k}}=[\lambda
k_{x}(k_{x}^{2}-3k_{y}^{2})+Jm_{z}]/(\hbar v_{F})$ and $\chi =\pm $ are the
upper and lower bands. Notice that the in-plane magnetization $m_{x/y}$ on
the dispersion cannot be eliminated by performing a gauge transformation due
to the existence of the warping term.

In linear response, we retain the non-equilibrium distribution function up
to the first order $f^{(1)}(\varepsilon _{\mathbf{k}}^{\chi })$ and
calculate the linear polarization $\mathbf{S}^{oc(1)}=\sum_{\chi }\frac{d^{d}%
\mathbf{k}}{(2\pi )^{d}}\mathbf{s}_{\chi }(\mathbf{k})f^{(1)}(\varepsilon _{%
\mathbf{k}}^{\chi })$. We find that $\mathbf{S}^{oc(1)}$ only contributes to
the field-like SOT (FL-SOT) and no antidamping SOT arises even for the case with
strong warping (see Supplement materials\cite{Supplemental Material}
or our previous work\cite{J.-Y.Li2019}). Here, extending the theory to the nonlinear one, we calculate the non-linear spin polarization with $\mathbf{S}^{oc(2)}\mathbf{=}\sum_{\chi }\frac{d^{d}\mathbf{k}}{(2\pi
)^{d}}\mathbf{s}_{\chi }(\mathbf{k}) f^{(2)}(\varepsilon _{\mathbf{k}}^{\chi
})$. We assume the Fermi level $\varepsilon _{F}>0$ lies in the upper surface-band $
\chi =1$ and the band index is suppressed afterwards. Choosing the in-plane
electric field $\mathbf{E}=(E_{x},E_{y})$, at low temperatures
we obtain the analytical expressions for the nonlinear spin polarization
\begin{eqnarray}
\begin{split}
S_{x}^{oc(2)} =&C\left[
a_{2}m_{x}E_{y}^{2}+\left( a_{1}m_{z}-a_{2}m_{y}\right) E_{x}E_{y}\right] ,
\\
S_{y}^{oc(2)} =&C\left[
 \left(\frac{a_{1}m_{z}}{2}+a_{2}m_{y}\right) E_{x}^{2} -\frac{a_{1}m_{z}}{2}E_{y}^{2}-a_{2}m_{x}E_{x}E_{y}%
\right] , \\
S_{z}^{oc(2)} =&C\left[\left( \frac{3}{2}a_{1}m_{y}-a_{0}m_{z}\right) E_{x}^{2}-\left( a_{0}m_{z}+\frac{3}{2}a_{1}m_{y}\right)
E_{y}^{2}\right] \\
&+3Ca_{1}m_{x}E_{x}E_{y},
\end{split}
\label{eq_S2}
\end{eqnarray}%
where we retain up to the second-order term of $\mathbf{m}$ and $\lambda$, and denote $C=e^{2}\gamma ^{2}v_{F}J/(8\pi)$, $a_{0}=1/(\hbar v_{F}\varepsilon _{F})$, $a_{1}=3\lambda \varepsilon _{F}/(\hbar^{4}v_{F}^{4})$, and $a_{2}=3\lambda ^{2}\varepsilon _{F}^{3}/(\hbar^{7}v_{F}^{7})$.

Interestingly, unlike the even function of $\mathbf{m}$ appearing in linear
response, the nonlinear spin polarizations in Eq. (\ref{eq_S2}) are odd functions
of $\mathbf{m}$ while all the second-order terms in $\mathbf{m}$ disappear.
Thus, the nonlinear spin polarization contributes an antidamping SOT $%
\mathbf{\tau }_{D}^{oc}=\frac{2J}{\hbar }\mathbf{m}\times \mathbf{S}^{oc(2)}%
$ with strength $\tau _{d}^{oc}=\frac{2J}{e\hbar |\mathbf{E}|}|\mathbf{S}%
^{oc(2)}|$. This result is unexpected since the change in the electron
occupation on the Fermi surface usually only contributes to the
FL-SOT. In FM/HM or FM/TI bilayer, the existing mechanisms for the
antidamping SOT include the contribution from the Berry curvature\cite%
{K.-S.Lee2016,H.Kurebayashi2014} or the electric-field-induced intrinsic
interband transition\cite%
{T.Chiba2017,A.Sakai2014,T.Yokoyama2010,J.-Y.Li2019,H.Li2015} or extrinsic
disorder-induced interband-coherence effects\cite{X.Cong2017}. There are
also some emerging new mechanisms such as interface spin currents\cite%
{V.P.Amin2018}, spin anomalous Hall effect\cite{S.Iihama2018}, nonreciprocal
generation of spin current\cite{G.Okano2019}, planar Hall current\cite%
{C.Safranski2019}, and magnon\cite{N.Okuma2017}. These mechanisms are based
on the linear response theory. Here, we propose an alternative mechanism
only associated with the intraband transitions beyond linear response
theory.

\begin{figure}[tbp]
\centering \includegraphics[width=0.48\textwidth]{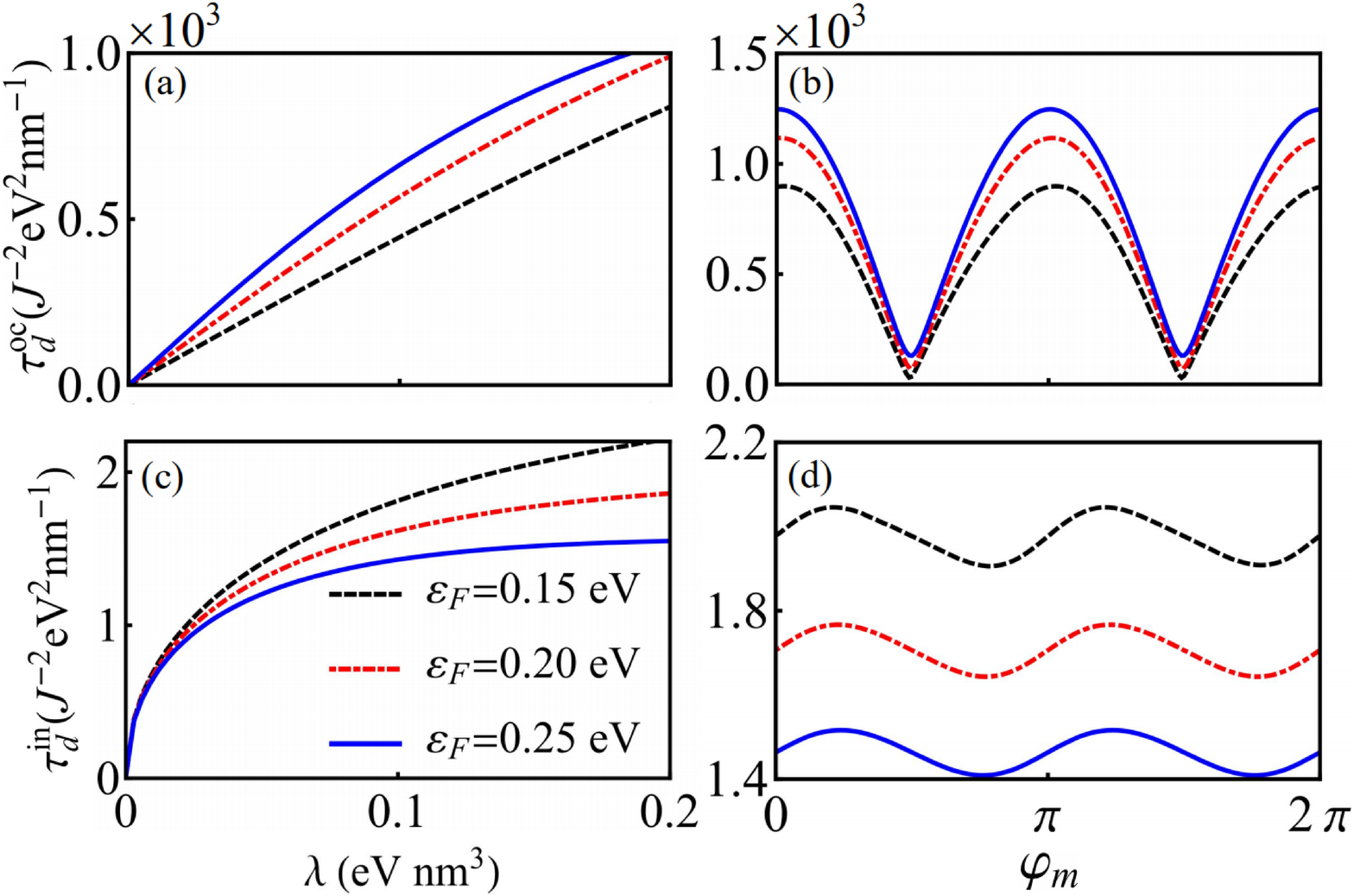}
\caption{Dependence of the strength of SOT (a) $\protect\tau _{d}^{oc}$ and (c) $\protect\tau _{d}^{in}$ on the warping parameter $\protect\lambda $ with constant azimuthal angle of $\mathbf{m}$ ($\protect\varphi_{m}=\pi/4$) for different Fermi energies. (b) and (d) The $\protect\varphi_{m}$-dependent strength of SOT with constant $\lambda=0.15$ ${\rm eV\cdot nm^{3}}$. Other parameters are set as: $\protect\theta _{m}=\protect\pi /2$, $\theta_{E} =\protect\pi /4$, $v_{F}=5\times 10^{5}$ ${\rm m/s}$, $\gamma =3 \rm ps$, and $|\mathbf{E}|=0.2{\rm mV/nm}$.}
\label{fig2}
\end{figure}

In order to illustrate the role of the warping effect, we set $\lambda =0,$ and Eq.
(\ref{eq_S2}) reduces to $S_{z}^{oc(2)}=\frac{-e^{2}\gamma ^{2}v_{F}J}{8\pi }%
a_{0}m_{z}|\mathbf{E}|^{2}$, which is controlled only by $m_{z}$ and
proportional to $1/\varepsilon _{F}$, and the other components vanish. This
implies that the linear-$\mathbf{k}$ Dirac dispersion also can give rise to
a non-linear spin polarization, which is distinct from the electric field-induced
nonlinear current\cite{P.He2018,vig} where the current $\mathbf{j}$ $\propto
\lambda $. Thus the current-spin correspondence fails in the non-linear regime
even in the absence of warping. For finite warping $%
\lambda \neq 0$, not only $m_{z}$ but also the in-plane magnetization $%
m_{x/y}$ play a role. Besides modifying the magnitude of $S_{z}^{oc(2)}$,
$m_{x/y}$ also generate extra in-plane components $S_{x/y}^{oc(2)}$. Importantly,
all of warping-related components are proportional to $\varepsilon _{F}$ and
$\lambda $ or their higher orders. We calculate the numerical
result of $\tau _{d}^{oc}$ directly with $\mathbf{S}^{oc(2)}\mathbf{=}\int\frac{d^{2}\mathbf{k}}{(2\pi
)^{2}}\mathbf{s}(\mathbf{k}) f^{(2)}(\mathbf{k})$ rather than with the analytical expressions Eq. (5), and present the numerical result of $\tau _{d}^{oc}$ as a function of $\lambda$ in Fig. \ref{fig2}(a), where all parameters are within the range of realistic TI materials. Prominently, the resulting ADL-SOT $\tau _{d}^{oc}$ strength increases remarkably as $\lambda $ or $\varepsilon _{F}$
increases, Therefore, for large $\varepsilon _{F}$ and $\lambda $, $\tau
_{d}^{oc}$ can be enhanced significantly in comparison with the case of Refs.\cite%
{T.Chiba2017,A.Sakai2014,T.Yokoyama2010,J.-Y.Li2019} in
absence of warping. In addition, $\tau
_{d}^{oc}$ exhibits a complex dependence on the azimuthal angle of $%
\mathbf{m}$, as shown in Fig. \ref{fig2}(b). A complex angular
dependence of the SOT has been observed in recent experimental measurements in TI
bilayers \cite{A.R.Mellnik2014,Y.Fan2014,M.DC2018,Y.Fan2016} but has not been explained theoretically to date.

\emph{Understanding of intraband nonlinear damping SOT}- For the linear Dirac
case ($\lambda =0$), a current-spin correspondence $\hat{v}=v_{F}%
\hat{z}\times \mathbf{\sigma }$ can be established due to spin-momentum locking, and then
nonequilibrium spin polarization reads $\mathbf{S}=-\frac{1}{ev_{F}}\hat{z}%
\times \mathbf{j}$. Here, the longitudinal conductance contributes to the
FL-SOT and the transversal conductance contributes to the damping-like SOT. This correspondence
relation is satisfied only for linear spin polarization without $\lambda $
and is broken by warping \cite{J.-Y.Li2019}. In the non-linear case,
even for $\lambda =0$, \ the correspondence\ is not followed since $%
S_{z}^{oc(2)}=\frac{e^{2}\gamma ^{2}v_{F}J}{8\pi }a_{0}m_{z}|\mathbf{E}|^{2}$
but $\mathbf{j}\propto $ $\lambda $ vanishes (see Section B of the supplementary material\cite{Supplemental Material}).
Thus, we cannot simply attribute the non-linear spin polarization to the
non-linear longitudinal or transverse conductance.

When applying an electric field on the TI surface, the hexagonal warped
Fermi surface would shift in $\mathbf{k}$-space, and generates a net
linear spin accumulation due to the spin-momentum locking, as given by the $%
\mathbf{m}$-independent term in Eq. (A.7) of supplementary materials\cite{Supplemental Material}. However,
this shift cannot generate a non-linear spin accumulations as given in Eq.
(5), where all terms are related to the magnetization $\mathbf{m}$. In order
to understand the generation of the nonlinear spin polarization, we need to
analyze the symmetry of the integrand in $\mathbf{S}^{oc(2)}=\int \frac{d^{2}%
\mathbf{k}}{(2\pi )^{2}}\mathbf{s}(\mathbf{k})f^{(2)}(\mathbf{k})$. In the
absence of the magnetization, the average spin $\mathbf{s}(\mathbf{k})$ is
odd in $\mathbf{k}$ whereas the second-order distribution function $ f^{(2)}(%
\mathbf{k})$ is even in $\mathbf{k}$. As a consequence, $\mathbf{S}^{oc(2)}=0$.

\begin{figure}[tbp]
\centering \includegraphics[width=0.3\textwidth]{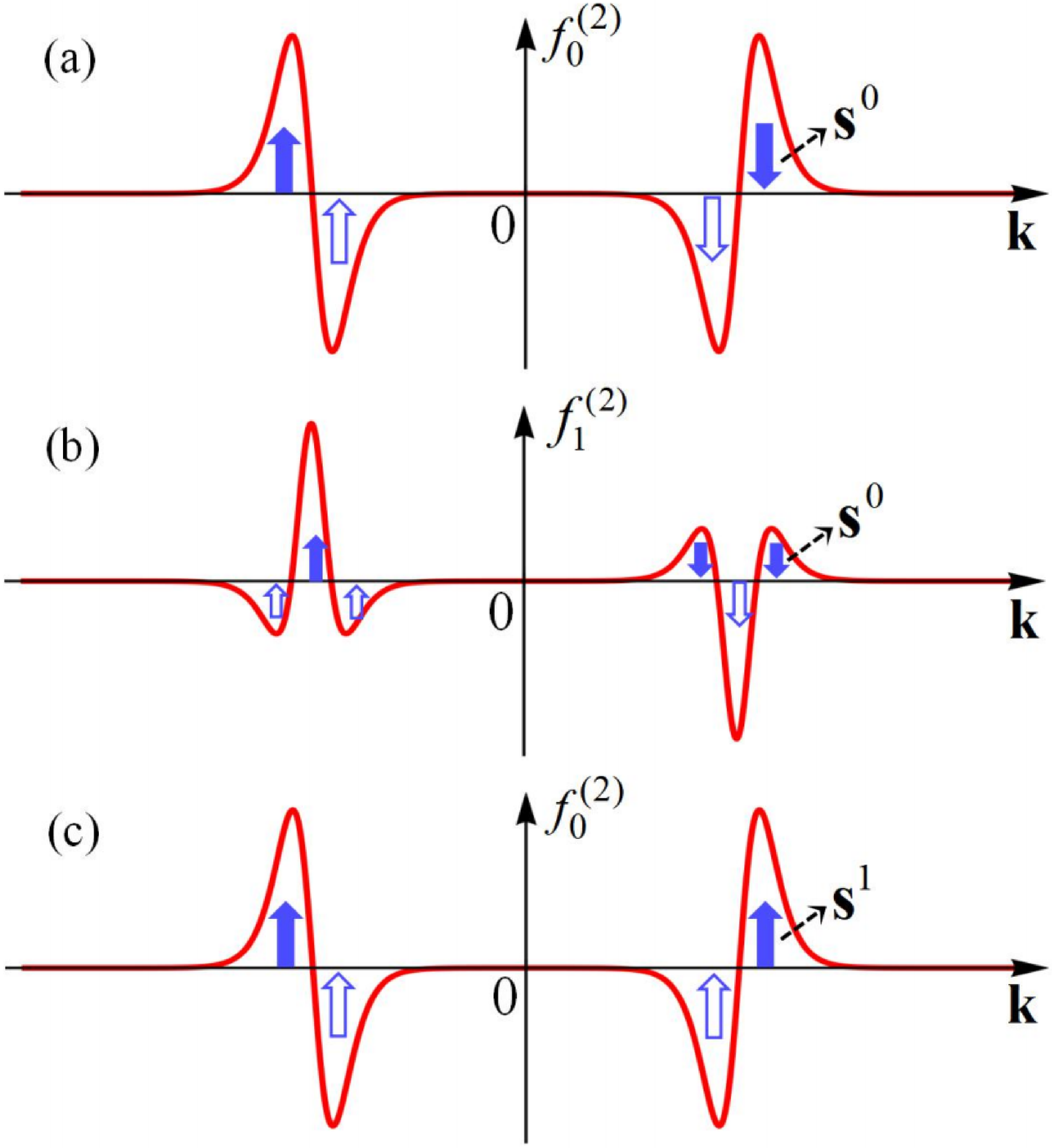}
\caption{Schematics of variation of the second-order correction $f_{i}^{(2)}$ of the distribution function along the $\mathbf{k}$-axis parallel to the applied
electric field $\mathbf{E}$. Blue solid
arrows represent excess of electrons with spins along the arrow direction, and hollow arrows represent depletion of the same. (a) describes the vanishing contribution from the component $\mathbf{s}^{0}(\mathbf{k}) \cdot f_{0}^{(2)}(\mathbf{k})$. The nonzero polarization stemming from the components (b) $\mathbf{s}^{0}(\mathbf{k}) \cdot f_{1}^{(2)}(\mathbf{k})$ and (c) $\mathbf{s}^{1}(\mathbf{k}) \cdot f_{0}^{(2)}(\mathbf{k})$.}
\label{fig3}
\end{figure}

When $\mathbf{m}$ is introduced, however, the warped Fermi surface is
further distorted except for the shift, which not only changes the
occupation of the electron states but also perturbs the spin textures, giving
an additional deviation to the spin direction at each $\mathbf{k}$-point. In
this case, both $\mathbf{s}(\mathbf{k})$ and $f^{(2)}(\mathbf{k})$ have symmetric and asymmetric components with respect to $\mathbf{m}$. Up to second-order in $J$ or $\mathbf{m}$ (see Section C of the supplementary materials\cite{Supplemental Material}), we expand $\mathbf{s}(\mathbf{k})=\sum_{i=0,1,2}\mathbf{s}^{i}(\mathbf{k})J^{i}$
and $f^{(2)}(\mathbf{k})=\sum_{i=0,1,2}f_{i}^{(2)}J^{i}.$ It is easy to
check that $\mathbf{s}^{i}(\mathbf{k})$ is odd and $f_{i}^{(2)}$ is even in $%
\mathbf{k}$ for $i=0,2$, while $\mathbf{s}^{i}(\mathbf{k})$ is even and $%
f_{i}^{(2)}$ is odd for $i=1$. Thus, the nonzero integrand terms of $\mathbf{%
k}$ in $\mathbf{S}^{oc(2)}$ are $\mathbf{s}^{0}(\mathbf{k}) \cdot f_{1}^{(2)}(%
\mathbf{k})$ and $\mathbf{s}^{1}(\mathbf{k}) \cdot f_{0}^{(2)}(\mathbf{k})$. In
Fig. 3, we plot the variation of the second-order correction $f_{i}^{(2)}$ of the distribution function along the $\mathbf{k}$-axis parallel to the applied
electric field $\mathbf{E}$. In the absence of $\mathbf{m}$ as in
Figs. \ref{fig3}(a), the occupations of the electron states at $\mathbf{k}$ and $-\mathbf{k}$ are the same but the corresponding spins are opposite, which contributes no net spin polarization. Once a nonzero $\mathbf{m}$ is introduced, the component $f_{1}^{(2)}(\mathbf{k})$ of the occupation or $\mathbf{s}^{1}(\mathbf{k})$ change parity. In Fig. 3(b) where both $f_{1}^{(2)}(\mathbf{k})$ and $\mathbf{s}^{0}(\mathbf{k})$ are odd, the down-spin electrons are depleted while the up-spin ones are in excess, which makes the opposite spins carried by the electrons in $\mathbf{k}$
and $-\mathbf{k}$ cannot cancel each other, and so a net nonlinear spin polarization appears for $\mathbf{s}^{0}(\mathbf{k}) \cdot f_{1}^{(2)}(%
\mathbf{k})$. Compared with Fig. 3(a), this is a result of $f^{(2)}(\mathbf{k})$ changing
from even to odd function, namely, $f_{0}^{(2)}(\mathbf{k})\rightarrow
f_{1}^{(2)}(\mathbf{k})$, by the interplay of the magnetization and the
nontrivial spin texture of warping effect. Fig. 3(c) describes the case of $\mathbf{s}^{1}(\mathbf{k}) \cdot f_{0}^{(2)}(\mathbf{k})$, where the even $f^{(2)}(\mathbf{k})$ keeps the same,  compared with Fig. 3(a), but $\mathbf{s}(\mathbf{k}%
)$ is changed from odd to even function $\mathbf{s}^{0}(\mathbf{k}%
)\rightarrow \mathbf{s}^{1}(\mathbf{k})$. $\mathbf{s}^{1}(-\mathbf{k})=%
\mathbf{s}^{1}(\mathbf{k})$ means the same spin orientations at $\mathbf{k}$
and $-\mathbf{k}$, which mainly originate from the deviation of out-of-plane
spin in warping effect by the magnetization or by out-of-plane $m_{z}$. The
latter contributes $S_{z}^{oc(2)}\propto a_{0}m_{z}$, which will quickly
shrink for the Fermi energy away from the Dirac point due to $a_{0}\propto
\frac{1}{\varepsilon _{F}}$. Physically, the distortion of Fermi surface
leads to change of spin texture and unequal population of electrons with
opposite momenta as well as spins and so generates the nonlinear spin
polarization. The increasing parameter $\lambda $ will enhance the
distortion effect and then the nonlinear spin polarization.

\emph{Comparison of nonlinear SOT with intrinsic SOT}- It is interesting
to compare the non-linear antidamping SOT with that from the Berry curvature
caused by intrinsic interband transitions, $\mathbf{S}^{in}\mathbf{=}%
\sum_{\chi }\frac{d^{d}\mathbf{k}}{(2\pi )^{d}}\delta \mathbf{s}_{\chi }(%
\mathbf{k})f(\varepsilon _{\mathbf{k}}^{\chi })$, where $\delta \mathbf{s}%
_{\chi }(\mathbf{k})=(\hbar/2) $Re$\langle \Psi _{\mathbf{k}}^{\chi }|%
\boldsymbol{\sigma }|\delta \Psi _{\mathbf{k}}^{\chi }\rangle $. By
modifying the quasiparticle wave functions $|\delta \Psi _{\mathbf{k}}^{\chi
}\rangle $ with perturbation method, the spin polarization is given by\cite%
{X.Cong2017,H.Kurebayashi2014,I.Garate2009},

\begin{equation}
\mathbf{S}^{in}=\frac{e\hbar^2 }{V}\sum\limits_{\chi \neq \chi ^{\prime
},\mathbf{k}}\left[f(\varepsilon _{\mathbf{k}}^{\chi })-f(\varepsilon _{\mathbf{k}%
}^{\chi ^{\prime }})\right]\frac{\mathrm{{Im}\left[ \left\langle \Psi _{\mathbf{k}%
}^{\chi }\right\vert \mathbf{\sigma}\left\vert \Psi _{\mathbf{k}}^{\chi ^{\prime
}}\right\rangle \left\langle \Psi _{\mathbf{k}}^{\chi ^{\prime }}\right\vert
\mathbf{\hat{v}}\cdot \mathbf{E}\left\vert \Psi _{\mathbf{k}}^{\chi
}\right\rangle \right] }}{\left(\varepsilon _{\mathbf{k}}^{\chi }-\varepsilon _{\mathbf{k%
}}^{\chi ^{\prime }}\right)^{2}}.  \label{eq_Sin}
\end{equation}%
This expression is analogous to the intrinsic Berry curvature mechanism
originally introduced to explain the anomalous Hall effect\cite%
{N.Nagaosa2010} and the SHE\cite{J.Sinova2014} due to the
electric-field-induced interband-coherence. It is found that this
antidamping Berry-curvature SOT can contribute with a strength comparable to
that of the SHE-driven antidamping torque, and has given good explanation for
ADL-SOT experiments with Rashba-model ferromagnets\cite{H.Kurebayashi2014}.

Here, we apply this intrinsic Berry curvature mechanism to the FM/TI
bilayer. For $\lambda =0$, we obtain $S_{z}^{in}=0$ and
\begin{equation}
S_{x/y}^{in}=\frac{e\hbar J}{8\pi }a_{0}m_{z}E_{x/y}.
\end{equation}%
Obviously, only $m_{z}$ contributes to the spin polarization and in turn the
intrinsic damping SOT, which recalls the results of Ref.\cite%
{I.Garate2010,P.B.Ndiaye2017,T.Chiba2017,A.Sakai2014,T.Yokoyama2010,T.Chiba2020}
based on the Green's function Kubo formula. For $\lambda \neq 0$, $m_{x/y}$
also play a role for the intrinsic damping. In view of the complex analytical expressions we only present the numerical results $\tau _{d}^{in}=\frac{2J}{e\hbar |\mathbf{E}|}| \mathbf{S}^{in}|$ in
Figs. \ref{fig2}(c) and (d).

We compare the non-linear SOT strength $\tau _{d}^{oc}$ in
Figs. \ref{fig2}(a)-(b) to the Berry-curvature SOT strength $\tau _{d}^{in}$
in Figs. \ref{fig2}(c)-(d). One can find that: (I) While $\tau _{d}^{in}$ slightly increases with the warping parameter $\lambda$, $\tau _{d}^{oc}$ increases quickly.
(II) As $\varepsilon _{F}$ increases, $\tau
_{d}^{oc}$ increases while $\tau _{d}^{in}$ decreases. Thus, Figure 2 shows that $\tau _{d}^{oc}$ is larger than $\tau _{d}^{in}$ by two to three orders of magnitude for chosen parameters, which are in the range of realistic materials. In practice, the applied electric field strength in FM/TI SOT experiments\cite{J.Han2017,Y.Wang2017,M.DC2018} is estimated as $|\mathbf{E}|=0.1$-$0.3$ $\rm mV/nm$, the relaxation time in the TI Bi$_2$Se$_3$ is typically\cite{gli} $\gamma =3$ ${\rm ps}$, and the warping parameter\cite{P.He2018,kuro} is $\lambda=0.056$-$0.18$ $\rm eV\cdot nm^{3}$. In addition, different from $\tau_{d}^{in}$, $\tau _{d}^{oc}$ shows the more complicate angular dependence on $\mathbf{m}$, compared Figs. \ref{fig2}(b) with (d).

\begin{figure}[h]
\centering \includegraphics[width=0.49
\textwidth]{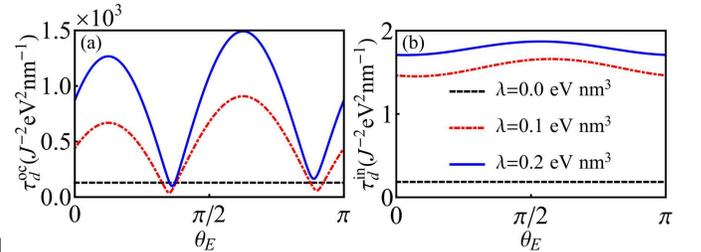}
\caption{The strength of SOT (a) $\protect\tau _{d}^{oc}$ and (b) $\protect\tau %
_{d}^{in}$ with respect to the current direction $\theta_E$ for different warping
parameters. Parameters are set as $\protect\varepsilon %
_{F}=0.2{\rm eV}$, $\protect\theta _{m}=%
\protect 19\pi /40$, and $\protect\varphi _{m}=\protect\pi /4.$. Other parameters are the same as in Fig. 2.}
\label{fig4}
\end{figure}

Owing to the warping effect, the current-induced SOT depends on the
current direction. In order to clarify the current-induced anisotropy of the
SOT, we plot $\tau _{d}^{oc}$ and $\tau _{d}^{in}$ as a function of the
direction of the electric field $\theta _{E}$ in Figs. \ref{fig4} (a) and (b),
respectively. Obviously, the ADL-SOTs are isotropic for $\lambda =0$ and anisotropic for $\lambda
\neq 0$, the larger the warping parameter (or Fermi energy), the more obvious the anisotropy is. More importantly, the periods of these two kinds of SOTs are significantly different, which could lead to an enhanced ratio $%
\tau _{d}^{oc}/\tau _{d}^{in}$ for a certain current direction $%
\theta _{E}$.

\emph{Conclusion-} We have studied the current-induced nonlinear spin polarization and SOT in a FM/TI bilayer with hexagonal warping. We focus on the single-band case by employing the Boltzmann equation, and find that the nonlinear spin polarization associated with intraband transitions generates a strong ADL-SOT, unlike the spin polarization linear in the electric field, which only contributes to the FL-SOT. The non-linear antidamping SOT not only stems from the out-plane magnetization $m_{z}$, but also from the joint effect of warping and in-plane magnetizations $m_{x}$ and $m_{y}$. The present mechanism is associated with intraband transitions, distinguished from the existing linear-response theory\cite{K.-S.Lee2016,H.Kurebayashi2014,T.Chiba2017,A.Sakai2014,T.Yokoyama2010,J.-Y.Li2019,H.Li2015,X.Cong2017}, where inter-band transitions are necessary. More importantly, the non-linear ADL-SOT is enhanced with increasing Fermi energy and warping parameter, and can be several orders of magnitude larger than that of the intrinsic Berry-curvature contributions. It exhibits a complex dependence on the azimuthal angle of the magnetization, which is consistent with experiment. This nonlinear SOT provides a new mechanism for the explanation of the giant SOT in recent experiments.

\acknowledgements This work was supported by GDUPS (2017), by the National
Natural Science Foundation of China (Grants No. 12174121, No. 11904107, and No. 12104167), by the Guangdong NSF of China (Grants No.
2021A1515010369 and No. 2020A1515011566), by Guangdong Basic and Applied
Basic Research Foundation (Grant No. 2020A1515111035). DC is supported by the Australian Research Council Future Fellowship FT190100062. LW is supported by the Australian Research Council Centre of Excellence in Future Low-Energy Electronics Technologies CE170100039.

\end{document}


\title{Supplementary material to\\
\textquotedblleft Non-linear antidamping spin-orbit torque originating from intra-band transport on the warped surface of a topological insulator\textquotedblright}
\maketitle

\section{\textbf{Derivation of the linear spin polarization}}

The $n$-th order nonequilibrium distribution function derived from the Eq.
(2) of the main text can be rewritten as
\begin{eqnarray}
f^{(1)}(\varepsilon _{\mathbf{k}}^{\chi }) &=&e\gamma \mathbf{E}\cdot
\mathbf{v}\frac{\partial f^{(0)}(\varepsilon _{\mathbf{k}}^{\chi })}{%
\partial \varepsilon _{\mathbf{k}}^{\chi }},  \label{Eq_f} \\
f^{(2)}(\varepsilon _{\mathbf{k}}^{\chi }) &=&\frac{e^{2}\gamma ^{2}}{\hbar }%
\left[ \mathbf{E}\cdot \frac{\partial (\mathbf{E}\cdot \mathbf{v})}{\partial
\mathbf{k}}\frac{\partial f^{(0)}(\varepsilon _{\mathbf{k}}^{\chi })}{%
\partial \varepsilon _{\mathbf{k}}^{\chi }}+\hbar (\mathbf{E}\cdot \mathbf{v}%
)^{2}\frac{\partial ^{2}f^{(0)}(\varepsilon _{\mathbf{k}}^{\chi })}{\partial
(\varepsilon _{\mathbf{k}}^{\chi })^{2}}\right] .
\end{eqnarray}%
where $\mathbf{v}=\frac{1}{\hslash }\partial _{\mathbf{k}}\varepsilon _{%
\mathbf{k}}^{\chi }$ is the group velocity of electrons and an in-plane
electric field $\mathbf{E}=(E_{x},E_{y})$ is applied. For convenience, we
choose a positive Fermi energy (i.e., $\varepsilon _{F}>0$ lies in the upper
band $\chi =1$). According to the eigenvalues $\varepsilon _{\mathbf{k}%
}^{\chi }$ shown in the Eq. (4) of the main text, the group velocity $%
\mathbf{v}=(v_{x},v_{y})$ can be easily solved as
\begin{equation}
\begin{split}
v_{x}& =\frac{1}{\hbar \varepsilon _{\mathbf{k}}^{+}}\left[ \hslash
v_{F}\left( \hbar v_{F}k_{x}-Jm_{y}\right) +3\lambda \left(
k_{x}^{2}-k_{y}^{2}\right) \Lambda _{\mathbf{k}}\right] , \\
v_{y}& =\frac{1}{\hbar \varepsilon _{\mathbf{k}}^{+}}\left[ \hbar
v_{F}\left( \hbar v_{F}k_{y}+Jm_{x}\right) -6\lambda k_{x}k_{y}\Lambda _{%
\mathbf{k}}\right] .
\end{split}%
\end{equation}

One can calculate the spin polarization using the following formula
\begin{equation}
\mathbf{S}^{oc}\mathbf{=}\sum_{\chi }\frac{d^{d}\mathbf{k}}{(2\pi )^{d}}%
\mathbf{s}_{\chi }(\mathbf{k})\delta f(\varepsilon _{\mathbf{k}}^{\chi }),
\end{equation}%
where $\delta f(\varepsilon _{\mathbf{k}}^{\chi })=f(\varepsilon _{\mathbf{k}%
}^{\chi })-f^{(0)}(\varepsilon _{\mathbf{k}}^{\chi })$ and $\mathbf{s}_{\chi
}(\mathbf{k})=(\hbar /2)\langle \Psi _{\mathbf{k}}^{\chi }|\boldsymbol{%
\sigma }|\Psi _{\mathbf{k}}^{\chi }\rangle $ is the spin expectation.
Diagonalizing the Hamiltonian of Eq. (3) of the main text, the corresponding
eigenstates can be solved as
\begin{equation}
\begin{split}
|\Psi _{\mathbf{k}}^{+}\rangle & =\left[ \cos \left( \xi /2\right) ,e^{i\eta
}\sin \left( \xi /2\right) \right] ^{T}, \\
|\Psi _{\mathbf{k}}^{-}\rangle & =\left[ -\sin \left( \xi /2\right)
,e^{i\eta }\cos \left( \xi /2\right) \right] ^{T},
\end{split}%
\end{equation}%
with $\cos (\xi )=\Lambda _{\mathbf{k}}/\varepsilon _{\mathbf{k}}^{+}$ and $%
\tan (\eta )=(Jm_{y}-\hbar v_{F}k_{x})/(Jm_{x}+\hbar v_{F}k_{y})$. Using the
above eigenstates, one can calculate the spin expectation $\mathbf{s}_{+}(%
\mathbf{k})=(s_{x},s_{y},s_{z})$ as
\begin{equation}
\mathbf{s}_{+}(\mathbf{k})=\left( \frac{J\hbar m_{x}+\hbar ^{2}v_{F}k_{y}}{%
2\varepsilon _{\mathbf{k}}^{+}},\frac{J\hbar m_{y}-\hbar ^{2}v_{F}k_{x}}{%
2\varepsilon _{\mathbf{k}}^{+}},\frac{\hbar \lambda k_{x}^{3}-3\hbar \lambda
k_{x}k_{y}^{2}+\hbar Jm_{z}}{2\varepsilon _{\mathbf{k}}^{+}}\right) .
\end{equation}

Substituting Eqs. (A.1) and (A.6) to the spin polarization $\mathbf{S}^{oc}$
of Eq. (A.4) and expanding the integrand of Eq. (A.4) to the second-order
term of $\mathbf{m}$ and $\lambda $, one can obtain the results for the
linear spin polarization $\mathbf{S}^{oc(1)}$, which reads
\begin{equation}
\begin{split}
S_{x}^{oc(1)}& =\frac{e\gamma }{8\pi }\left[ \frac{\varepsilon _{F}}{\hbar
v_{F}}-\frac{1}{6}a_{2}\varepsilon _{F}^{2}-a_{0}J^{2}m_{z}^{2}-\frac{3}{2}%
a_{2}J^{2}(m_{x}^{2}+m_{y}^{2}-m_{z}^{2})\right] E_{y}, \\
S_{y}^{oc(1)}& =\frac{e\gamma }{8\pi }\left[ -\frac{\varepsilon _{F}}{\hbar
v_{F}}+\frac{1}{6}a_{2}\varepsilon _{F}^{2}+a_{0}J^{2}m_{z}^{2}+\frac{3}{2}%
a_{2}J^{2}(m_{x}^{2}+m_{y}^{2}-m_{z}^{2})\right] E_{x}, \\
S_{z}^{oc(1)}& =\frac{e\gamma J^{2}}{8\pi }\left\{ \left[
a_{1}(-m_{x}^{2}+m_{y}^{2})-3a_{2}m_{y}m_{z}\right]
E_{x}+(2a_{1}m_{x}m_{y}+3a_{2}m_{x}m_{z})E_{y}\right\} ,
\end{split}%
\end{equation}%
where $a_{0}=1/(\hbar v_{F}\varepsilon _{F})$, $a_{1}=3\lambda \varepsilon
_{F}/(\hbar ^{4}v_{F}^{4})$, and $a_{2}=3\lambda ^{2}\varepsilon
_{F}^{3}/(\hbar ^{7}v_{F}^{7})$. As shown above, all components of the
linear spin polarization are even function of $\mathbf{m}$. Thus, the linear
spin polarization $\mathbf{S}^{oc(1)}$ only contributes to the field-like
SOT.

\section{\textbf{Derivation of the nonlinear current}}

For a positive Fermi energy, the charge current density can be calculated by
\begin{equation}
\mathbf{j}=-e\int \frac{d^{d}\mathbf{k}}{(2\pi )^{d}}\mathbf{v}%
_{k}f(\varepsilon _{\mathbf{k}}^{+}).  \label{Eq_j}
\end{equation}%
Considering the linear response theory, i.e., substituting Eqs. (A.1) and
(A.3) into the above equation, the linear current density $\mathbf{j}^{(1)}$
in FM/TI can be easily solved after some algebraic calculations. The
resulting $\mathbf{j}^{(1)}$ reads as
\begin{equation}
\begin{split}
\mathbf{j}^{(1)}& =\sigma _{D}\mathbf{E}, \\
& =\frac{e^{2}\gamma \varepsilon _{F}}{4\pi \hslash ^{2}}\mathbf{E}.
\end{split}%
\end{equation}%
For the nonlinear response, one can substitute the non-equilibrium
distribution function $f_{k,\chi }^{(2)}$ of Eq. (A.2) to Eq. (B.1).
Following the similar algebraic calculations, the nonlinear current density $%
\mathbf{j}^{(2)}=(j_{x}^{(2)},j_{y}^{(2)})$ can be obtain
\begin{equation}
\begin{split}
j_{x}^{(2)}& =c_{1}\left(
2m_{x}E_{x}E_{y}-3m_{y}E_{x}^{2}-m_{y}E_{y}^{2}\right)
-c_{2}m_{z}(E_{x}^{2}-E_{y}^{2}), \\
j_{y}^{(2)}& =c_{1}\left(
m_{x}E_{x}^{2}-2m_{y}E_{x}E_{y}+3m_{x}E_{y}^{2}\right)
+2c_{2}m_{z}E_{x}E_{y},
\end{split}%
\end{equation}%
where $c_{1}=3e^{3}\gamma ^{2}\lambda ^{2}J\varepsilon _{F}^{3}/(4\pi \hbar
^{8}v_{F}^{5}),c_{2}=3e^{3}\gamma ^{2}\lambda J\varepsilon _{F}/(8\pi \hbar
^{5}v_{F}^{2})$. As shown above, $\mathbf{j}^{(2)}=0$ when $\lambda
\rightarrow 0$.

\section{\textbf{Derivation of the nonlinear spin polarization}}

In this section, we discuss in detail the generation of the nonlinear spin
polarization. To facilitate the analysis, we need to expand $\mathbf{s}(%
\mathbf{k})$ and $f^{(2)}(\varepsilon _{\mathbf{k}}^{+})$ to the
second-order term of $\mathbf{m}$ (or the equivalent $J$), i.e., $\mathbf{s}(%
\mathbf{k})=\sum_{i=0,1,2}\mathbf{s}^{i}(\mathbf{k})J^{i}$ and $%
f^{(2)}(\varepsilon _{\mathbf{k}}^{\chi })=\sum_{i=0,1,2}f_{i}^{(2)}(\mathbf{%
k})J^{i}$.

For the convenience of discussions, we here simply set the electric field as
$\mathbf{E}=(E_{x},0)$. For arbitrary direction of electric field, the case
is similar. In the weak warping limit, $\mathbf{s}^{i}(\mathbf{k})$ and $%
f_{i}^{(2)}(\mathbf{k})$ can be expand to the second-order term of $\lambda $%
, which is enough to capture the warping effect. In this way, one can obtain
the analytical expressions of $\mathbf{s}^{i}$ and $f_{i}^{(2)}$, $\mathbf{s}%
(\mathbf{k})$ reads as%
\begin{equation*}
\begin{split}
\mathbf{s}^{0}(\mathbf{k})& =\left( \frac{\hbar ^{2}v_{F}k_{y}}{2\varepsilon
_{0}},\;-\frac{\hbar ^{2}v_{F}k_{x}}{2\varepsilon _{0}},\;\frac{\hbar
\lambda k_{x}^{3}-3\hbar \lambda k_{x}k_{y}^{2}}{2\varepsilon _{0}}\right) ,
\\
\mathbf{s}^{1}(\mathbf{k})& =\left( \frac{\hbar \varepsilon _{0}m_{x}-\hbar
^{2}v_{F}k_{y}\kappa _{3}}{2\varepsilon _{0}^{2}},\;\frac{\hbar \varepsilon
_{0}m_{y}+\hbar ^{2}v_{F}k_{x}\kappa _{3}}{2\varepsilon _{0}^{2}},\;\frac{%
\hbar \varepsilon _{0}m_{z}-\hbar \lambda k_{x}^{3}\kappa _{3}+3\hbar
\lambda k_{x}k_{y}^{2}\kappa _{3}}{2\varepsilon _{0}^{2}}\right) , \\
\mathbf{s}^{2}(\mathbf{k})& =\left[ \frac{\hbar ^{2}v_{F}k_{y}(3\kappa
_{3}^{2}-\mathbf{m}^{2})-2\hbar \varepsilon _{0}m_{x}\kappa _{3}}{%
2\varepsilon _{0}^{3}},\;\frac{-\hbar ^{2}v_{F}k_{x}(3\kappa _{3}^{2}-%
\mathbf{m}^{2})+2\hbar \varepsilon _{\mathbf{0}}m_{y}\kappa _{3}}{%
2\varepsilon _{0}^{3}},\;\frac{\hbar \lambda
(k_{x}^{3}-3k_{x}k_{y}^{2})(3\kappa _{3}^{2}-\mathbf{m}^{2})-2\hbar
\varepsilon _{0}m_{z}\kappa _{3}}{2\varepsilon _{0}^{3}}\right] ,
\end{split}%
\end{equation*}%
where $\varepsilon _{0}=\sqrt{\lambda
^{2}(k_{x}^{3}-3k_{x}k_{y}^{2})^{2}+(\hbar v_{F}k)^{2}}$ and $\kappa _{3}=%
\left[ (k_{x}^{3}-3k_{x}k_{y}^{2})\lambda m_{z}+\hbar
v_{F}(k_{y}m_{x}-k_{x}m_{y})\right] /\varepsilon _{0}$. It is found that $%
\mathbf{s}^{0,2}(\mathbf{k})$ is odd in $\mathbf{k}$ while $\mathbf{s}^{1}(%
\mathbf{k})$ is even.

$f_{i}^{(2)}$ reads as%
\begin{equation*}
\begin{split}
f_{0}^{(2)}& =\frac{e^{2}\gamma ^{2}E_{x}^{2}}{\hbar }\left[ \frac{\kappa
_{4}-\hbar ^{2}\kappa _{5}^{2}}{\hbar \varepsilon _{0}}\frac{%
-be^{b(\varepsilon _{0}-\varepsilon _{F})}}{(1+e^{b(\varepsilon
_{0}-\varepsilon _{F})})^{2}}+\hbar \kappa _{5}^{2}\frac{b^{2}e^{b(%
\varepsilon _{0}-\varepsilon _{F})}\left[ e^{b(\varepsilon _{0}-\varepsilon
_{F})}-1\right] }{\left[ 1+e^{b(\varepsilon _{0}-\varepsilon _{F})}\right]
^{3}}\right] , \\
f_{1}^{(2)}& =\frac{e^{2}\gamma ^{2}E_{x}^{2}}{\hbar }\left\{ \frac{3\hbar
\kappa _{3}\kappa _{5}^{2}-\hbar \kappa _{3}\kappa _{4}-2\hbar \kappa
_{5}\left( 3\lambda (k_{x}^{2}-k_{y}^{2})m_{z}-\hbar v_{F}m_{y}\right)
+6\lambda k_{x}\varepsilon _{0}m_{z}}{\hbar \varepsilon _{0}^{2}}\times
\frac{-be^{b(\varepsilon _{0}-\varepsilon _{F})}}{(1+e^{b(\varepsilon
_{0}-\varepsilon _{F})})^{2}}\right. \\
& +\frac{\kappa _{4}-\hslash ^{2}\kappa _{5}^{2}}{\hbar \varepsilon _{0}}%
\times \frac{b^{2}e^{b(\varepsilon _{0}-\varepsilon _{F})}\left[
-1+e^{b(\varepsilon _{0}-\varepsilon _{F})}\right] \kappa _{3}}{\left[
1+e^{b(\varepsilon _{0}-\varepsilon _{F})}\right] ^{3}} \\
& +\frac{2\kappa _{5}\left[ 3\lambda (k_{x}^{2}-k_{y}^{2})m_{z}-\hbar
v_{F}m_{y}\right] -2\hbar \kappa _{3}\kappa _{5}^{2}}{\varepsilon _{0}}%
\times \frac{b^{2}e^{b(\varepsilon _{0}-\varepsilon _{F})}\left[
e^{b(\varepsilon _{0}-\varepsilon _{F})}-1\right] }{\left[
1+e^{b(\varepsilon _{0}-\varepsilon _{F})}\right] ^{3}} \\
& -\left. \hbar \kappa _{5}^{2}\times \frac{b^{3}e^{b(\varepsilon
_{0}-\varepsilon _{F})}\left[ 1-4e^{b(\varepsilon _{0}-\varepsilon
_{F})}+e^{2b(\varepsilon _{0}-\varepsilon _{F})}\right] \kappa _{3}}{\left[
1+e^{b(\varepsilon _{0}-\varepsilon _{F})}\right] ^{4}}\right\} , \\
f_{2}^{(2)}& =\frac{e^{2}\gamma ^{2}E_{x}^{2}}{\hbar }\left\{ \frac{3\hbar
^{2}(5\kappa _{3}^{2}+\mathbf{m}^{2})\kappa _{5}^{2}+12\hbar ^{2}\kappa
_{3}\kappa _{5}\left[ 3\lambda (k_{x}^{2}-k_{y}^{2})m_{z}-\hbar v_{F}m_{y}%
\right] -2\left[ 3\lambda (k_{x}^{2}-k_{y}^{2})m_{z}-\hbar v_{F}m_{y}\right]
^{2}+3\kappa _{3}^{2}\kappa _{4}\varepsilon _{0}}{\hbar \varepsilon _{0}^{3}}%
\right. \\
& \frac{-be^{b(\varepsilon _{0}-\varepsilon _{F})}}{(1+e^{b(\varepsilon
_{0}-\varepsilon _{F})})^{2}}+2\frac{3\hbar \kappa _{3}\kappa _{5}^{2}-\hbar
\kappa _{3}\kappa _{4}-2\hbar \kappa _{5}\left[ 3\lambda
(k_{x}^{2}-k_{y}^{2})m_{z}-\hbar v_{F}m_{y}\right] +6\lambda
k_{x}\varepsilon _{0}m_{z}}{\hbar \varepsilon _{0}^{2}}\times \frac{\kappa
_{3}b^{2}e^{b(\varepsilon _{0}-\varepsilon _{F})}\left[ -1+e^{b(\varepsilon
_{0}-\varepsilon _{F})}\right] }{\left[ 1+e^{b(\varepsilon _{0}-\varepsilon
_{F})}\right] ^{3}} \\
& +\frac{\kappa _{4}-\hbar ^{2}\kappa _{5}^{2}}{\hbar \varepsilon _{0}}\left[
\frac{b^{2}e^{b(\varepsilon _{0}-\varepsilon _{F})}\left[ 1-e^{b(\varepsilon
_{0}-\varepsilon _{F})}\right] (\kappa _{3}^{2}-\mathbf{m}^{2})}{\left[
1+e^{b(\varepsilon _{0}-\varepsilon _{F})}\right] ^{3}\varepsilon _{0}}-%
\frac{\kappa _{3}^{2}b^{3}e^{b(\varepsilon _{0}-\varepsilon _{F})}\left[
1-4e^{b(\varepsilon _{0}-\varepsilon _{F})}+e^{2b(\varepsilon
_{0}-\varepsilon _{F})}\right] }{\left[ 1+e^{b(\varepsilon _{0}-\varepsilon
_{F})}\right] ^{4}}\right] \\
& +\hbar \left[ \frac{2(4\kappa _{3}^{2}-\mathbf{m}^{2})\kappa _{5}^{2}}{%
\varepsilon _{0}^{2}}-\frac{8\kappa _{3}\kappa _{5}\left[ 3\lambda
(k_{x}^{2}-k_{y}^{2})m_{z}-\hbar v_{F}m_{y}\right] }{\hbar \varepsilon
_{0}^{2}}+\frac{2\left[ 3\lambda (k_{x}^{2}-k_{y}^{2})m_{z}-\hbar v_{F}m_{y}%
\right] ^{2}}{\hbar ^{2}\varepsilon _{0}^{2}}\right] \frac{%
b^{2}e^{b(\varepsilon _{0}-\varepsilon _{F})}\left[ e^{b(\varepsilon
_{0}-\varepsilon _{F})}-1\right] }{\left[ 1+e^{b(\varepsilon
_{0}-\varepsilon _{F})}\right] ^{3}} \\
& +2\frac{2\kappa _{5}\left[ 3\lambda (k_{x}^{2}-k_{y}^{2})m_{z}-\hbar
v_{F}m_{y}\right] -2\hbar \kappa _{3}\kappa _{5}^{2}}{\varepsilon _{0}}%
\times \frac{\kappa _{3}b^{3}e^{b(\varepsilon _{0}-\varepsilon _{F})}\left[
1-4e^{b(\varepsilon _{0}-\varepsilon _{F})}+e^{2b(\varepsilon
_{0}-\varepsilon _{F})}\right] }{\left[ 1+e^{b(\varepsilon _{0}-\varepsilon
_{F})}\right] ^{4}} \\
& +\hbar \kappa _{5}^{2}\left[ \frac{\kappa _{3}^{2}b^{4}e^{b(\varepsilon
_{0}-\varepsilon _{F})}\left[ -1+11e^{b(\varepsilon _{0}-\varepsilon
_{F})}-11e^{2b(\varepsilon _{0}-\varepsilon _{F})}+e^{3b(\varepsilon
_{0}-\varepsilon _{F})}\right] }{\left[ 1+e^{b(\varepsilon _{0}-\varepsilon
_{F})}\right] ^{5}}\right. \\
& \left. \left. +\frac{b^{3}e^{b(\varepsilon _{0}-\varepsilon _{F})}\left[
1-4e^{b(\varepsilon _{0}-\varepsilon _{F})}+e^{2b(\varepsilon
_{0}-\varepsilon _{F})}\right] (\kappa _{3}^{2}-\mathbf{m}^{2})}{\left[
1+e^{b(\varepsilon _{0}-\varepsilon _{F})}\right] ^{4}\varepsilon _{0}}%
\right] \right\} ,
\end{split}%
\end{equation*}%
where $\kappa _{4}=(15k_{x}^{4}-36k_{x}^{2}k_{y}^{2}+9k_{y}^{4})\lambda
^{2}+(\hbar v_{F})^{2}$ is even in $\mathbf{k}$ and $\kappa _{5}=\left[
(3k_{x}^{5}-12k_{x}^{3}k_{y}^{2}+9k_{x}k_{y}^{4})\lambda ^{2}+(\hbar
v_{F})^{2}k_{x}\right] /(\hbar \varepsilon _{0})$ is odd in $\mathbf{k}$,
and $b=1/(k_{B}T)$. It is found that $f_{0,2}^{(2)}(\mathbf{k})$ is even in $%
\mathbf{k}$ while $f_{1}^{(2)}(\mathbf{k})$ is odd in $\mathbf{k}$.

All the above results indicate that the nonzero integrand terms of $\mathbf{k%
}$ in $\mathbf{S}^{oc(2)}$ are $\mathbf{s}^{0}(\mathbf{k}) \cdot f_{1}^{(2)}(%
\mathbf{k})$ and $\mathbf{s}^{1}(\mathbf{k}) \cdot f_{0}^{(2)}(\mathbf{k})$.